\documentclass[twocolumn,showpacs,preprintnumbers,amsmath,amssymb]{revtex4}

\usepackage{graphicx}
\usepackage{dcolumn}
\usepackage{bm}

\def\btbl{\begin{tabular}} \def\etbl{\end{tabular}}
\def\CFM{{\footnotesize CFM}} \def\TCM{{\footnotesize TCM}}
 
 \def\SPS{{\footnotesize SPS}}
\def\E877{{\footnotesize E877}} \def\QGP{{\footnotesize QGP}}
\def\E941{{\footnotesize E941}} \def\E864{{\footnotesize E864}}
\def\NA49{{\footnotesize NA49}} \def\NA35{{\footnotesize NA35}}
\def\RHIC{{\footnotesize RHIC}} \def\PHOBOS{{\footnotesize PHOBOS}}
\def\HCM{{\footnotesize HCM}} 
\def\NA{{\footnotesize NA}} 
 \def\AGS{{\footnotesize AGS}}
\def\RHIC{{\footnotesize RHIC}} \def\LHC{{\footnotesize LHC}}

\def\AMPT{{\footnotesize AMPT}}

\begin{document}
\title{ Thermalization component model of multiplicity distributions of
    charged hadrons measured at the
BNL($E^{lab}_{NN}$=2-11.6GeV), the CERN( $E^{lab}_{NN}$=20-200GeV)
, and the BNL ($\sqrt{s_{NN}}$=19.6-200GeV) }

\author{Shengqin ~Feng$^{1,2}$}
\author{Wei ~Xiong$^1$}

\affiliation{$^1$College of Science, China Three Gorges Univ.,
Yichang 443002, Hubei China }
\affiliation{$^2$School of Physics
and Technology, Wuhan University ,Wuhan 430072, Hubei, China}

\begin{abstract}
We find that collective flow model which can successfully analyze
charged particle distributions at AGS and lower SPS
($E^{lab}_{NN}$ less than 20GeV in the lab frame). but fails to
analyze that of at RHIC. The tails of distribution of charged
particle at RHIC has a jump from the collective flow model
calculation as the energy increases. Thermalization Component
Model is presented based on collective flow to study the
multiplicity distributions at RHIC in this paper. It is realized
that the region of phase space of collective flow can reflect that
of thermalization region. By comparing the contributions of
particle productions from thermalization region at different
energies and different centralities, we can deepen
our study on the feature of collective movement at RHIC.\\
\vskip0.2cm \noindent Keywowds: ~~Thermalization,~~Thermalization
component model
\end{abstract}

\pacs{25.75.Ld, 25.75.Dw} \maketitle

\section{Introduction}
\label{intro} One of the central question at \RHIC~ is the extent
to which the quanta produced in the collision interact and
thermalize~\cite{Jacobs,Mueller}. Nuclear collisions generate
enormous multiplicity and transverse energy, but in what extent
does the collision generate matter in local equilibrium which can
be characterized by the thermodynamic parameters temperature,
pressure, and energy density?  Only if thermalization has been
established can more detailed questions be asked about the
equation of state of the matter.

Recently it is realized that the study of collective flow is one
of the important tools to study multi-hadron production of
relativistic heavy-ion collisions~\cite{Mayer,Braun,Blume}. This
is because the longitudinal and transverse flow includes rich
physics, and collective flow relates closely to early evolution
and nuclear stopping.  Collective flow is often utilized to
express the thermalization degree of relativistic heavy-ion
collisions system. Detailed studies of the observed final state
flow pattern will deep our understanding of dynamic mechanism of
relativistic heavy-ion collisions. {\em Collective Flow Model}
~~\cite{Schnedermann1,Schnedermann2} (\CFM) was developed basing
on the pure thermal model. It achieves success in the discussion
of charged particle distribution at \AGS~ and lower \SPS~~ energy
(20GeV) and become an indicator for the existence of collective
flow at \AGS\cite{Schnedermann1,Schnedermann2}. But a detailed
analysis of the experimental data at \SPS~~ and \RHIC~~ energy
with \CFM~ has shown that as the increase of collision energy, the
two tails of the charged hadron distributions have a symmetric
jump away from the calculation of \CFM. These phenomena will make
us to reconsider collective flow theory at higher collision
energy.

As shown in Fig.1, \CFM~ fails to analyze the charged hadron
distribution at \RHIC~ energy regions. As the increase of the
collision energy, the tail of the distribution of experimental
data jump from the calculation of \CFM. The naive reason seems to
be that experimental data on hadron yields are now available over
a broad collision energy range as the increase of collision
energy.  As the increase of phase space of particle distribution,
thermalization at whole phase space of particle production becomes
more difficult. Detailed analysis of thermalization relation with
centrality and energies at \RHIC~ is needed.

\begin{figure}[h!]
\centering \resizebox{0.45\textwidth}{!}{
\includegraphics{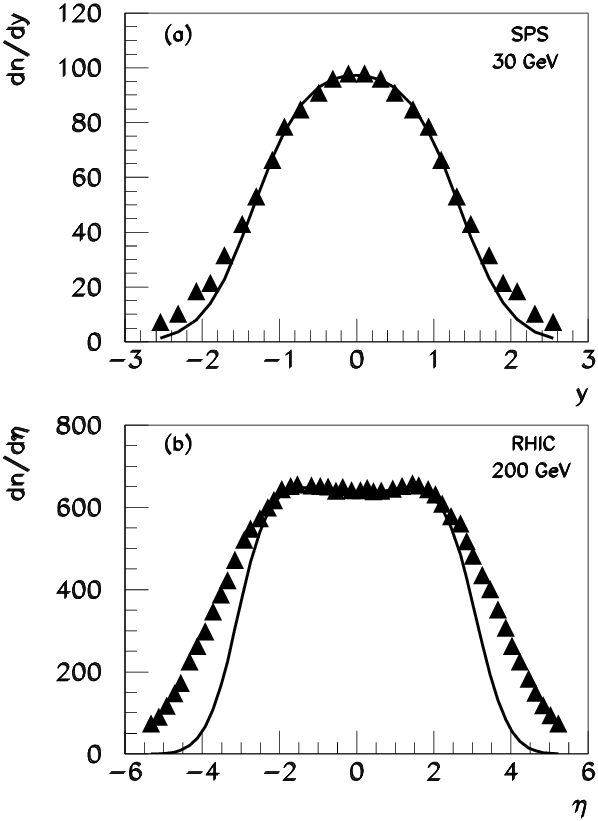}}
\caption{(a) $\pi$ meson rapidity distribution of
$E^{lab}_{NN}$=30 GeV at \SPS~~\cite{Blume}; (b) charged hadron
pseudo-rapidity distribution of $\sqrt{s_{NN}}$=200 GeV}
\label{fig1}
\end{figure}

\begin{figure}[h!]
\centering \resizebox{0.48\textwidth}{!}{
\includegraphics{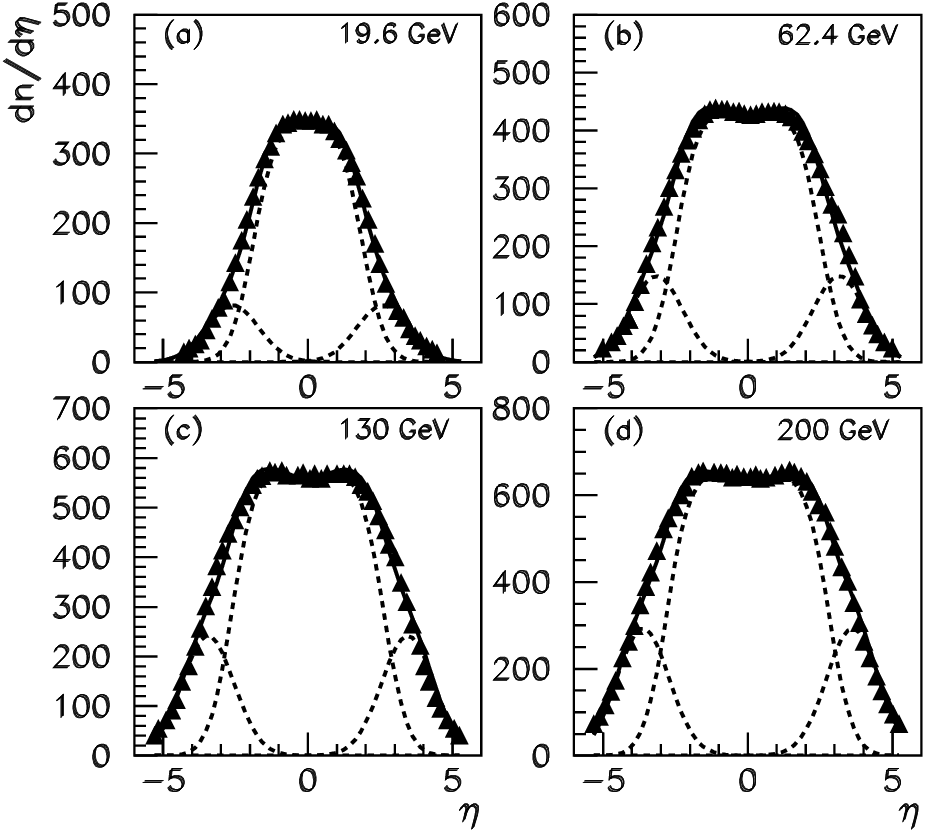}}
\caption{The pseudo-rapidity distributions at $\sqrt{s_{NN}}$
=19.6, 62.4, 130, 200 GeV for Au+Au Collisions. Experimental data
are given by triangle~~\cite{Back1,Back2,Back3}.The solid lines
are the results given by \TCM, which is the summation of the three
component contributions.} \label{fig2}
\end{figure}

How to simulate the data of charged hadron distribution at higher
\SPS~ and \RHIC~ energy regions and what these results tell us are
the main topic of the paper.  \PHOBOS~ has used three Gaussian
distributions to simulate the distribution of charged hadron
successfully~\cite{Back1,Back2,Back3,Alexandru}.  Georg
Wolschin~~\cite{Wolschin} et al also discussed the charged hadron
distribution by using three component distribution functions to
construct Fokker£­Plank equation. Their models both assumed three
random Gaussian distribution emitting sources. The Three Gaussian
sources represent target, projectile and central source in
physics, respectively.

The main goal of this paper is to give a study of the
thermalization features of multi-particle production in heavy ion
collision at high energy in the framework of the collective flow
theory. We restrict ourselves here to the basic features and
essential results of the \CFM~ approach.  A complete survey of the
assumptions and results, as well as of the relevant references, is
available in Ref.\cite{Schnedermann1,Schnedermann2},
\cite{Feng1}-\cite{Yuan}.

The paper is organized as follows. The analysis details based on
the {\em Thermalization Component Model}(\TCM) are described in
Sec.2. The comparisons of \TCM~ calculations with experimental
data and the related theoretical analysis with \TCM are given in
Sec.3. A summary is given in Sec. 4.

\section{Thermalization component model}
The hot and dense matter produced in relativistic heavy ion
collisions may evolve through the following scenario:
pre-equilibrium, thermal (or chemical) equilibrium of partons,
possible formation \QGP~ or a \QGP~ hadron gas mixed state, a gas
of hot interacting hadrons, and finally, a freeze-out state when
the produced hadrons no longer strongly interact with each other.
Since the produced hadrons carry information about the collision
dynamics and the entire space-time evolution of the system from
the initial to the final stage of collisions, a precise analysis
of the multiplicity distributions of charged hadrons is essential
for the understanding of the dynamics and properties of the
created matter.

A detailed analysis of the experimental data at \SPS~ and \RHIC~
energy with \CFM~ has shown that as the increase of collision
energy,the two tails of $\pi$ or the charged hadron distributions
show a (symmetric) discrepancy between the data and the
calculation. These phenomena will make us to reconsider Collective
flow theory at higher collision energy. Detailed analysis of the
relation with thermalization with centralities and energies at
\RHIC~ is needed. Let us first sketch our overall picture and
detail our arguments subsequently. The model we considered
contains three distinct assumptions some of which are rather
different from those usually contained in other flow models.

(i) The size of phase space of the particle distribution increases
with the increase of collision energy.  It seems more difficult to
realize thermalization at the whole phase space of particle
production at \SPS~ and \RHIC~ data. It is assumed that the
Gaussian distributions were fit to the distributions of the
produced charged hadrons at the two fragmentation regions, and
thermalization prefers to occur at the central rapidity region at
\SPS~ and \RHIC.

(ii) The collective flow of central rapidity region carries
information of the early time of heavy-ion collision. The system
expands not only in the longitudinal direction, but also in the
transverse direction. The two dimensional collective flow is used
to study the thermalization process at \RHIC.

(iii) The phase space is compartmentalized as the thermalization
region and non-thermalization regions. The non-thermalization
regions locate at the two fragmentation regions. The total
multiplicity distributions are the summation of the contributions
from the target fragmentation region, projectile fragmentation
region and central region, respectively.

\begin{equation}  
\frac {dN}{dy}=N_{1}F_{1}+N_{2}F_{2}+N_{3}F_{3}=\sum_{i}N_{i}F_{i}
\label{eq:eq1} 
\end{equation}

Here $i=1,2,3$  denotes target, projectile and central region,
respectively. $N_{i}$  and $F_{i}$  are the particle numbers and
the normalization functions of target, projectile and central
regions, respectively.

As assumed before, the distributions of target and projectile
fragmentation regions are given with Gaussian distributions:

\begin{equation}  
F_{1}={\frac
{1}{\sqrt{2\pi}\sigma}}e^{-\frac{(y+y_{1})^{2}}{2\sigma^{2}}}
\label{eq:eq2} 
\end{equation}

\begin{equation}  
F_{2}=\frac
{1}{\sqrt{2\pi}\sigma}e^{-\frac{(y+y_{2})^{2}}{2\sigma^{2}}}
\label{eq:eq3} 
\end{equation}

\noindent Here $\sigma$  is the distribution width of Gaussian,
$y_{1},y_{2}$ are the locations of central of target and
projectile emitting source.

$F_{3}$ is the distribution of two dimensional flow, which is
given by~\cite{Schnedermann1},\cite{Schnedermann2}

\begin{eqnarray}  
F_{3}=&&\frac {g\tau_{f}R_{f}^{2}K}{8\pi}  \int_{m_{t}^{lo}}^{m_{t}^{hi}} dm_{t}^{2}m_{t}I_{0}(\alpha)\nonumber \\
     && \int_{-\eta_{0}}^{-\eta_{0}} d\eta_{l}\cosh(y-\eta_{l})e^{\mu/T}e^{{-\bar{\alpha}}\cosh(y-\eta_{l})}
\label{eq:eq4} 
\end{eqnarray}

Here $m_{t}^{lo}$   and $m_{t}^{hi}$ are the experimental limits
in which the spectrum is measured. The freeze-out radius $R_{f}$
and the longitudinal extend of the fireball is fixed via the
finite interval (-$\eta_{0}$, $\eta_{0}$), $I_{0}$is modified
Bessel function.

For the two dimensional flow theories, we should say a few words.
The geometry of the freeze-out of two dimensional flow
hyper-surface $\sigma_{f}$ fixed as follows: in the time direction
we take a surface of constant proper time . In $\eta_{l}$
direction, the freeze-out volume extends only to a maximum
space-time rapidity $\eta_{0}$ , which is required by the finite
available total energy and breaks longitudinal boost-invariance
proposed by Bjorken~~\cite{Bjorken}. In the transverse direction
the boundary is given by $R_{f}$, which describes a cylindrical
fireball in the $\eta-r$  space. The detailed discussion was shown
in Ref.\cite{Schnedermann1,Schnedermann2}.

\section{Comparing with Experimental data}

It is found that \CFM~ describe experimental data of charged
particle distribution very well when we discuss Au-Au center
collisions at \AGS~ energy region. The contribution of
fragmentation regions can be ignored, so expressions (1) can be
predigested:

\begin{equation}  
\frac{dN}{dy}=N_{3}F_{3}
\label{eq:eq5} 
\end{equation}

The results from \CFM~  are consistent with experimental data in
Au-Au collisions at \AGS~ energy region, such as $E^{lab}_{NN}$=2,
4, 6, 8, 11.6 GeV in the lab frame. This indicates that when at
lower \AGS~ energy region \CFM~ can describe the charged particle
distribution well, then our thermalization component model revert
to collective flow model. The reason seems that phase space is
small and the nucleus stopping power is very strong at \AGS~
energy region, so particles can be almost completely thermalized
in whole phase space. The same situation is true for that of \SPS~
energy region below 20 GeV.

But with the collision energies increase ($E^{lab}_{NN}$ above 30
GeV), the experimental points have a symmetric jump away from the
calculation of \CFM~ at two tails (as shown in Fig.1). This
phenomenon can be explained by the nuclei's penetrability. The
higher the collision energies, the more transparent the nuclei,
and the larger the extension of the phase space of the produced
particle. Collective flow is formed at the central rapidity region
after thermalization. The distributions of non-thermalization
charged hadrons are presented by Gaussian. The thermalization area
becomes one part of the whole phase space.

Since June 2000, the Relativistic Heavy-Ion Collider (\RHIC) has
opened a new energy region for the study of multi-hardon
production. We have analyzed the experimental data of charged
particle distribution in Au-Au center collisions in the  \RHIC~
energy region from 19.6 to 200 GeV of $\sqrt{s_{NN}}$

We can calculate the rapidity distribution of charged particles
whit expressions (1). It is known that we transfer rapidity
distribution to pseudo-rapidity distribution just by multiplying a
factor~~\cite{Wong}:

\begin{equation}  
\frac{dN}{d\eta}=\frac{dN}{dy}\sqrt{1-(\frac{m}{m_{T}\cosh{y}})^{2}}
\label{eq:eq6} 
\end{equation}

\begin{figure}[h!]
\centering \resizebox{0.46\textwidth}{!}{
\includegraphics{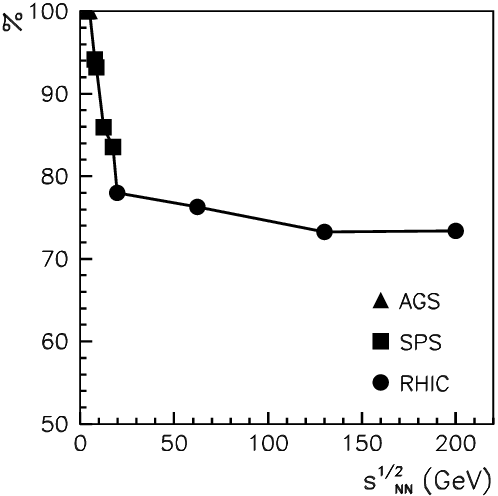}}
\caption{The dependence of the percentage of the charged hadron
production from the thermalization regions on the collision
energies.} \label{fig3}
\end{figure}

\begin{figure}[h!]
\centering \resizebox{0.48\textwidth}{!}{
\includegraphics{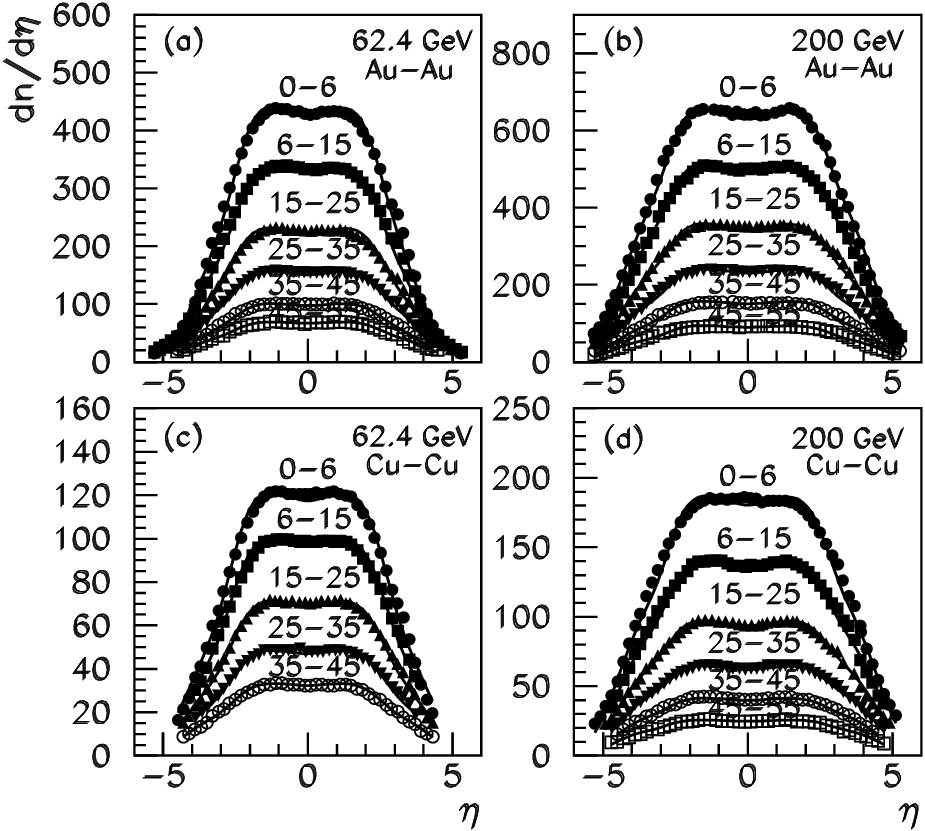}}
\caption{The charged hadron pseudo-rapidity distribution at
different centrality at $\sqrt{s_{NN}}$=62.4GeV and 200GeV for
Au+Au and Cu+Cu Collisions, respectively. Solid lines are the
results from \TCM, Experimental data are given by
\PHOBOS~~\cite{Back1}-\cite{Alexandru}} \label{fig4}
\end{figure}

We fit the experimental data of \RHIC~ energy region by \HCM~
model by $\chi^{2}/dof$ . The comparison of the measured and
calculated distributions for the best fit ( $\chi^{2}/dof$
minimization) is presented in Fig. 2. The \TCM~ calculations are
accordant with experimental data shown in Fig.2. The percentages
of the charged hadron productions from the thermalization regions
at \AGS, \SPS~ and \RHIC~ energy region are presented in Fig.3 by
\TCM.  It is found that most of the produced particles at \AGS~
come from the thermalization region, and the percentage of
produced particles from the thermalization region decreases as the
energy increase. The reduction trend becomes weaker and seems to
reach saturation as $\sqrt{s_{NN}}$ reaches 62.4GeV at \RHIC~
energy region.  The detailed fit parameters of our \TCM~ with
experimental data are shown in Table 1.

\begin{table}[h]
\caption{The fit results of \TCM~ with the experimental data at
\SPS~ and \RHIC~ energy regions}

\centering
\btbl{|c|c|c|c|c|c|c|}\hline
 &$E^{lab}_{NN}$&$\eta_{0}$&$y_{1,2}$&$n_{1}+n_{2}$&$n_{3}$&$n_{3}/(n_{1}+n_{2}+n_{3})$ \\ \hline
 &30&1.33&$\pm2.1$&16&256&94.13\% \\ \cline{2-7}
 &40&1.4&$\pm2.05$&22&301&93.19\% \\ \cline{2-7}
 &80&1.4&$\pm2.0$&64&392&85.97\% \\ \cline{2-7}
 SPS &158&1.38&$\pm2.0$&100&507&83.52\%  \\ \hline
&$\sqrt{s_{NN}}$&$\eta_{0}$&$y_{1,2}$&$n_{1}+n_{2}$&$n_{3}$&$n_{3}/(n_{1}+n_{2}+n_{3})$
\\ \hline
&19.6&1.85&$\pm2.6$&370&1310&77.99\% \\ \cline{2-7}
 &62.4&2.47&$\pm3.15$&670&2157&76.30\% \\ \cline{2-7}
 &130&2.62&$\pm3.45$&1100&3016&73.28\% \\ \cline{2-7}
 RHIC &200&2.8&$\pm3.62$&1320&3629&73.38\%  \\ \hline
\etbl
\end{table}

\PHOBOS~ Collaboration Working at \RHIC has presented many
experimental data~~\cite{Back1,Back2,Back3} of different energy
and different centrality including Au-Au collisions and Cu-Cu
collisions at $\sqrt{s_{NN}}$ =62.4 and 200 GeV. It is found that
the calculation results from \TCM~ are consistent with that of the
experimental data. The results are presented by Fig.4 and Table 2.
The experimental data are taken from
Ref.~\cite{Back1}-\cite{Alexandru}.

\begin{figure}[h!]
\centering \resizebox{0.48\textwidth}{!} {
\includegraphics{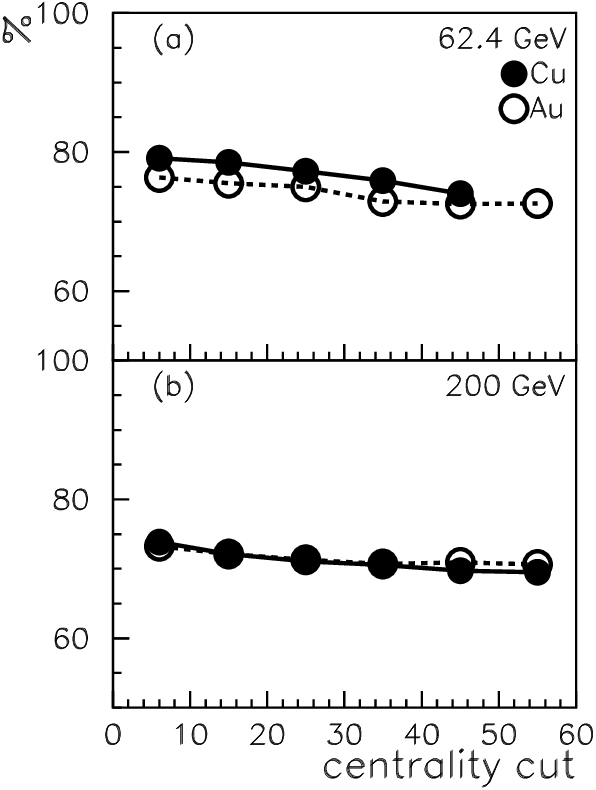}}
\caption{The dependence of the percentage from the thermalization
region on different centralities for $\sqrt{s_{NN}}$ =62.4, 200
GeV} \label{fig5}
\end{figure}

It is shown from Fig.5 that the percentage ratios of the particle
production from the thermalization regions increase with the
increase of the centralities at \RHIC. From Fig.5 (a), It is found
that the contribution ratios from the thermalization region is
appreciably larger for the smaller collision system (Cu +Cu) than
that of larger collision system (Au +Au) at $\sqrt{s_{NN}}$=
62.4GeV. But from Fig.5 (b), we find that the percentage ratios of
particle production from thermalization regions is almost
independent of the size of collision systems at
$\sqrt{s_{NN}}$=200 GeV .

In our \TCM , the free parameters are the limitation of collective
flow $\eta_{0}$  and the emission sources' positions in
fragmentation area $y_{1,2}$. We have $y_{1}=-y_{2}$  in the case
of symmetry collisions. The values of transverse flow and
temperature of collective flow refer to
Ref.\cite{Schnedermann1,Schnedermann2,Feng1,Yuan}. The values of
$n_{i}(i=1,2,3)$ are numbers of particles from the fragmentation
and the thermalization  regions. respectively.

A linear relationship is obtained between $\eta_{0}$ and
$ln\sqrt{s_{NN}}$  by detailed study. The linear equations are
given by fitting four data at \SPS~, \RHIC~ energy regions as
follows:

\begin{equation}  
\eta_{0}=0.40ln\sqrt{{s}_{NN}}+0.71
\label{eq:eq7} 
\end{equation}

here $\eta_{0}$  is the extension of collective flow.  From Eq.7,
we can predict the extension of the thermalization region at \LHC~
with the collision energy increase.

\begin{figure}[h!]
\centering \resizebox{0.48\textwidth}{!} {
\includegraphics{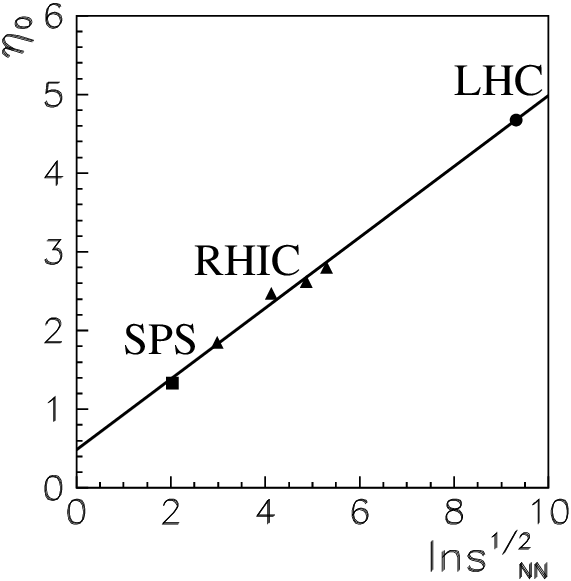}}
\caption{The relation between the limitation of thermalization
region with $ln\sqrt{s_{NN}}$ .} \label{fig6}
\end{figure}

Here, we should mention that quite a few theoretical models can
give equally good representation of the data of particle
productions at \AGS,\SPS~ and \RHIC, such as these thermal models
\cite{Braun1,Braun2,Becattini,Cleymans} based on the assumption of
global thermal and chemical equilibrium, and hydrodynamic models
\cite{Rischke1,Rischke2,Hung,Huovinen,Kolb1,Kolb2} based only on
the assumption of local thermal equilibrium, to transport models
\cite{Sorge1,Bass1,Molnar,Kahana,Li,Sa} that treat nonequilibrium
dynamics explicitly. 

The hydrodynamic models are particularly useful for
understanding the collective behavior of low transverse momentum
particles such as the elliptic flow, while the thermal models have been very successful
in accounting for the yield of various particles and their ratios. The transport models
are also natural and powerful tools for studying the Hanbury-Brown-Twiss
interferometry of hadrons Since they treat
chemical and thermal freeze-out dynamically.

Using parton distribution functions in the colliding nuclei, one \cite{Wang1} can study hard processes that involve large momentum
transfer based on the perturbative quantum chromodynamics (pQCD). Kharzeev et al \cite{Kharzeev1,Kharzeev2,Kharzeev3} developed the classical Yang-Mills
theory  to study  the evolution of parton distribution
functions in nuclei at ultra-relativistic energies and used to
study the hadron rapidity distribution and its centrality
dependence at \RHIC. These
problems have also been investigated in the pQCD-based final-state
saturation model \cite{Eskola1,Eskola2,Eskola3}. A multiphase
transport (\AMPT) model \cite{Lin1,Lin2,Lin3,Lin4} that includes both initial partonic and
final hadronic interactions and the transition between these two
phases of matter  was constructed to
describe nuclear interactions ranging from p - A to A - A systems at
center-of-mass energies from $\sqrt{s_{NN}}$ = 5 to 5500 GeV
at \LHC.

\section{Summary and conclusions}
Hadron multiplicities and their distributions are observables
which can provide information on the nature, composition, and size
of the medium from which they are originating. Of particular
interest is the extent to which the measured particle yields are
showing thermalization. The feature of thermalization of high
energy heavy ion collisions at \RHIC~ has been analyzed in this
paper.

\CFM~ fails to analyze the charged particle distributions when the
collision energies increase to above 30 GeV. The tail of
distribution of the charged particle at \RHIC~ has a jump from the
\CFM~ calculation with the energy increase. The naive reason seems
to be that the experimental data on hadron yields are now
available over a broad collision energy range with the increase of
collision energy.  It seems more difficult for thermalization at
the whole phase space of particle production with the increase of
the phase space of particle distribution.

On the other hand, the phenomena may suggest that something else
happens, including interaction mechanism, such as the onset of
de-confinement in the early stage of the reaction with the
collision energy ($E^{lab}_{NN}$)above 30 GeV at the lab frame,
which has been mentioned in Ref.\cite{Afanasiev}. In
Ref.\cite{Afanasiev}, central Pb - Pb collisions were studied in
the \SPS~ energy range. At around $E^{lab}_{NN}$=30 GeV the ratio
of strangeness to pion production shows a sharp maximum, the rate
of increase of the produced pion multiplicity per wounded nucleon
increases and the effective temperature of pions and kaons levels
to a constant value. These features are not reproduced by present
hadronic models, however there is a natural explanation in a
reaction scenario with the onset of de-confinement in the early
stage of the reaction at \SPS~ energy.

Collective flow in heavy-ion collisions is an unavoidable
consequence of thermalization. The extension of the phase space of
collective flow can reflect that of thermalization region.  It is
found that the \TCM~ can fit the experimental data well for the
particle production at the whole \AGS, \SPS~ and \RHIC~ energy
regions.  The percentage ratios of contributions of the particle
production from the thermalization region are the largest at \AGS,
and decrease as collision energies increase at \SPS~ and \RHIC,
but seem to reach saturation when $\sqrt{s_{NN}}$=62.4- 200 GeV at
\RHIC. It is also found that the extension of the flow shows a
linear dependence on $ln\sqrt{s_{NN}}$. From that, we can predict
the thermalization extension at future \LHC~ experimental data.

It is shown from our study that the percentage ratios of particle
production from thermalization regions increase with the increase
of the centralities at \RHIC. The contribution ratios from
thermalization region are appreciably larger for the smaller
collision system (Cu + Cu) at $\sqrt{s_{NN}}$=62.4GeV, but
independent of the collision system at $\sqrt{s_{NN}}$=200 GeV .

\section{Acknowledgments}
This work was supported by the Excellent Youth Foundation of Hubei
Scientific Committee(2006ABB036) and Natural Science Foundation of
China Three Gorges University (2003C02). The authors is indebted
to Prof. Lianshou Liu for his valuable discussions and very
helpful suggestions.

{}

\end{document}